\documentclass{aa}
\usepackage{hyperref}
\pdfoutput=1 
\usepackage{natbib}
\usepackage{txfonts}
\usepackage{graphicx}
\usepackage{indentfirst}
\usepackage{booktabs}
\usepackage{textcomp, gensymb}
\usepackage{hyperref}
\usepackage{ulem}

\makeatletter
\renewcommand*\aa@pageof{, page \thepage{} of \pageref*{LastPage}}

\makeatother

\begin{document} 
\renewcommand{\sectionautorefname}{Section}
\renewcommand{\subsectionautorefname}{Section}

 %%%%%%%%%%%%%%%%%%%%%%%%%%%%%%%%%%%%%%%%%%%%%%%%%%%%%%%%%%%%%%%%%%%%%%%%%%%%%%%%%%%%%%%%%
 %							TITLE
 %%%%%%%%%%%%%%%%%%%%%%%%%%%%%%%%%%%%%%%%%%%%%%%%%%%%%%%%%%%%%%%%%%%%%%%%%%%%%%%%%%%%%%%%%

\title{LOFAR imaging of the solar corona during the 2015 March 20 solar eclipse}

 %%%%%%%%%%%%%%%%%%%%%%%%%%%%%%%%%%%%%%%%%%%%%%%%%%%%%%%%%%%%%%%%%%%%%%%%%%%%%%%%%%%%%%%%%
 %							AUTHORS
 %%%%%%%%%%%%%%%%%%%%%%%%%%%%%%%%%%%%%%%%%%%%%%%%%%%%%%%%%%%%%%%%%%%%%%%%%%%%%%%%%%%%%%%%%

\author{A. M. Ryan\inst{1}\inst{2}\inst{3}, P. T. Gallagher\inst{2}, E. P. Carley\inst{2}, M. A. Brentjens\inst{4}, P. C. Murphy\inst{1}\inst{2}, C. Vocks\inst{5}, D. E. Morosan\inst{6}, H. Reid\inst{7}, J. Magdalenic\inst{8,9}, F. Breitling\inst{5}, P. Zucca\inst{4}, R. Fallows\inst{4}, G. Mann\inst{5}, A. Kerdraon\inst{10} \and R. Halfwerk\inst{3}}

\institute{
School of Physics, Trinity College Dublin, Dublin 2, Ireland.
\and
School of Cosmic Physics, Dublin Institute for Advanced Studies, D02 XF86, Ireland.
\and
AstroTec Holding B.V., Oude Hoogeveensedijk 4, 7991 PD Dwingeloo, Netherlands.
\and
ASTRON, The Netherlands Institute for Radio Astronomy, Oude Hoogeveensedijk 4, 7991 PD Dwingeloo, The Netherlands.
\and
Leibniz-Institut für Astrophysik Potsdam (AIP), An der Sternwarte 16, D-14482 Potsdam, Germany.
\and
Department of Physics, University of Helsinki, P.O. Box 64, FI-00014 Helsinki, Finland.
\and
Department of Space and Climate Physics, University College London, London, RH5 6NT, UK.
\and
Solar-Terrestrial Centre of Excellence—SIDC, Royal Observatory of Belgium, 3 Avenue Circulaire, B-1180 Uccle, Belgium.
\and
Center for mathematical Plasma Astrophysics, Department of Mathematics, KU Leuven, Celestijnenlaan 200B, B-3001 Leuven, Belgium
\and
LESIA, Observatoire de Paris, Université PSL, CNRS, Sorbonne Université, Universit de Paris, 5 Place Jules Janssen, 92195 Meudon, France.\\\\\email{ryana38@tcd.ie}}

% \abstract{}{}{}{}{} 
% 5 {} token are mandatory

 %%%%%%%%%%%%%%%%%%%%%%%%%%%%%%%%%%%%%%%%%%%%%%%%%%%%%%%%%%%%%%%%%%%%%%%%%%%%%%%%%%%%%%%%%
 %							ABSTRACT
 %%%%%%%%%%%%%%%%%%%%%%%%%%%%%%%%%%%%%%%%%%%%%%%%%%%%%%%%%%%%%%%%%%%%%%%%%%%%%%%%%%%%%%%%%

\abstract{The solar corona is a highly-structured plasma which can reach temperatures of more than $\sim$2~MK. At low frequencies (decimetric and metric wavelengths), scattering and refraction of electromagnetic waves are thought to considerably increase the imaged radio source sizes (up to a few arcminutes). However, exactly how source size relates to scattering due to turbulence is still subject to investigation. The theoretical predictions relating source broadening to propagation effects have not been fully confirmed by observations due to the rarity of high spatial resolution observations of the solar corona at low frequencies. Here, the LOw Frequency ARray (LOFAR) was used to observe the solar corona at 120--180~MHz using baselines of up to $\sim$3.5~km (corresponding to a resolution of $\sim$1--2\arcmin) during the partial solar eclipse of 2015 March 20. A lunar de-occultation technique was used to achieve higher spatial resolution ($\sim$0.6\arcmin) than that attainable via standard interferometric imaging ($\sim$2.4\arcmin). This provides a means of studying the contribution of scattering to apparent source size broadening. It was found that the de-occultation technique reveals a more structured quiet corona that is not resolved from standard imaging, implying scattering may be overestimated in this region when using standard imaging techniques. However, an active region source was measured to be $\sim$4\arcmin\space using both de-occultation and standard imaging. This may be explained by the increased scattering of radio waves by turbulent density fluctuations in active regions, which is more severe than in the quiet Sun.}


\titlerunning{Imaging the Solar Corona during the 2015 March 20 Eclipse using LOFAR}
\authorrunning{A. M. Ryan et al.}
\maketitle
 %%%%%%%%%%%%%%%%%%%%%%%%%%%%%%%%%%%%%%%%%%%%%%%%%%%%%%%%%%%%%%%%%%%%%%%%%%%%%%%%%%%%%%%%%
 %							INTRODUCTION
 %%%%%%%%%%%%%%%%%%%%%%%%%%%%%%%%%%%%%%%%%%%%%%%%%%%%%%%%%%%%%%%%%%%%%%%%%%%%%%%%%%%%%%%%%
\section{Introduction}\label{sec:intro} 
Solar radio observations are an invaluable tool to better understand solar eruptive processes and the structure of the solar corona. As radio waves travel through the corona, they are subject to propagation effects such as scattering off of electrons as well as refraction due to a changing electron density and refractive index \citep{erickson1964radio}. This results in apparent angular broadening of the radio sources, which is directly related to coronal density turbulence and inhomogeneities \citep{steinberg1971coronal}. The effect of apparent angular broadening is observable at decimetric and metric wavelengths and becomes more severe as the observed frequency approaches the local plasma frequency and if there are increasing levels of density fluctuations due to higher turbulence, for example, near an active region \citep{abramenko2010intermittency, abramenko2020analysis}. Therefore, the study of radio source size variation can provide greater insight into radio wave propagation effects as well as the nature of density inhomogeneities in the corona. However, to date, the exact relationship between source size and coronal scattering remains somewhat inconclusive. This ambiguity emphasises the importance and necessity of high spatial resolution, low frequency (10--300~MHz) radio observations of the smallest sources of compact radio emission in the corona.

Many observational studies of the solar corona at low frequencies have measured observed source sizes. Work done by \cite{lang1987vla} and \cite{zlobec1992vla} resolved sources of 30-–40\arcsec\space using the Very Large Array \cite[VLA;][]{thompson1980very, napier1983very} between 328--333MHz. A sub-arcminute structure associated with a Type I noise storm was observed by \cite{kerdraon1979observation} at 169 MHz using the Nançay Radioheliograph \cite[NRH;][]{kerdraon1997nanccay} and by \cite{mercier2006combining, mercier2015structure} at 236 MHz and 327 MHz using the NRH and the Giant Meterwave Radio Telescope \citep[GMRT;][]{ananthakrishnan2002multi} in tandem. At lower frequencies, \cite{2000A&A...358..749R} observed sources of approximately 3\arcmin\space at 34.5 MHz using the Decameter Wave Radio Telescope in the Gauribidanur Observatory \citep{sastry1995decameter}. However, due to insufficient baselines lengths, none of the above studies have imaged sub-arcsecond structure in the metric and decimetric regime, despite X-ray and extreme ultraviolet (EUV) imaging showing the corona to be highly structured on arcsecond and sub-arcsecond scales \citep{walker1988soft, koutchmy1988small, golub1990sub}. 

The lack of small-scale coronal structures at low frequencies is theoretically due to the large amount of scattering experienced by radio emission in coronal plasma. There have been numerous studies on the effects of scattering on the broadening of source sizes \citep{steinberg1971coronal}, the shift in source position \citep{fokker1965coronal}, and change in the intensity of observed radio emission \citep{riddle1974observation, robinson1983scattering}. A number of studies have carried out comparisons of observed source size to theoretical predictions of turbulence made with the use of coronal scattering models \citep{mcmullin1977effects, melrose1988implications, mercier2006combining, thejappa2008effects, subramanian2011constraints}. It has been theorised that the angular size of sources in solar radio observations is limited to arcminute scales due to this coronal scattering \citep{bastian1992scattering, bastian1994angular}. Recently, \citet{kontar2017imaging} showed scattering to be quite severe at low frequencies, using tied-array imaging to show that a 0.1\arcmin\space radio source, observed at 32\,MHz, can be broadened to $\sim$20\arcmin\space through scattering alone. However, the large size of the observed source may be due (in part) to the tied-array technique rather than inherent source size \citep{murphy2021lofar}.

Recently, the increasing need for improved resolution and sensitivity at low radio frequencies has encouraged the use of larger arrays spread across several hundreds of kilometres, such as the LOw Frequency ARray \cite[LOFAR;][]{van2013lofar} and on a smaller scale the Murchison Widefield Array \cite[MWA;][]{tingay2013murchison}. These arrays are now providing regular imaging of both the quiet and active Sun \citep{breitling2015lofar, mccauley2017type, vocks2018lofar, zhang2020interferometric}. The large baselines help in increasing the resolution of these instruments, allowing them to provide radio observations in the metric range with which we can probe the small-scale coronal structures. As well as this, Very Long Baseline Interferometry (VLBI) observations have been used to achieve sub-arcsecond resolution in the microwave regime \citep{tapping1983vlbi, benz1996very}. However, in the absence of longer baseline or indeed VLBI observations, radio solar eclipse observations can be exploited to achieve superior angular resolution. These high resolution observations can then be used to measure source sizes and further constrain the extent of scattering effects. This technique has been used with microwave observations, for example, \cite{marsh1980vla} and \cite{gary1987multifrequency}, where the motion of the lunar disc across the Sun provided the ability to resolve source sizes smaller than that possible using standard interferometry. This motivates a similar type of study in the metric range, where scattering is considered to be more prominent.

In this paper, interferometric LOFAR observations of a solar eclipse on 2015 March 20 are presented. This is the first LOFAR observation of a solar eclipse, which granted a unique opportunity to probe coronal source sizes via the lunar de-occultation technique. \autoref{sec:obs} gives context to the solar activity at the time of the observation and introduces the LOFAR telescope, providing a description of the instrument's specifications. Following on, an overview of the observing campaign is given. \autoref{sec:data_analysis} details the methods used for imaging and source size determination. \autoref{sec:results} focuses on the results of this work. Lastly, \autoref{sec:discussion} provides an analysis of the results in the context of previous observations.

%%%%%%%%%%%%%%%%%%%%%%%%%%%%%%%%%%%%%%%%%%%%%%%%%%%%%%%%%%%%%%%%%%%%%%%%%%%%%%%%%%%%%%%%%
%							OBSERVATIONS
%%%%%%%%%%%%%%%%%%%%%%%%%%%%%%%%%%%%%%%%%%%%%%%%%%%%%%%%%%%%%%%%%%%%%%%%%%%%%%%%%%%%%%%%%

\section{Observations}\label{sec:obs} 
\begin{figure}
  \centering
  \includegraphics[width=\columnwidth]{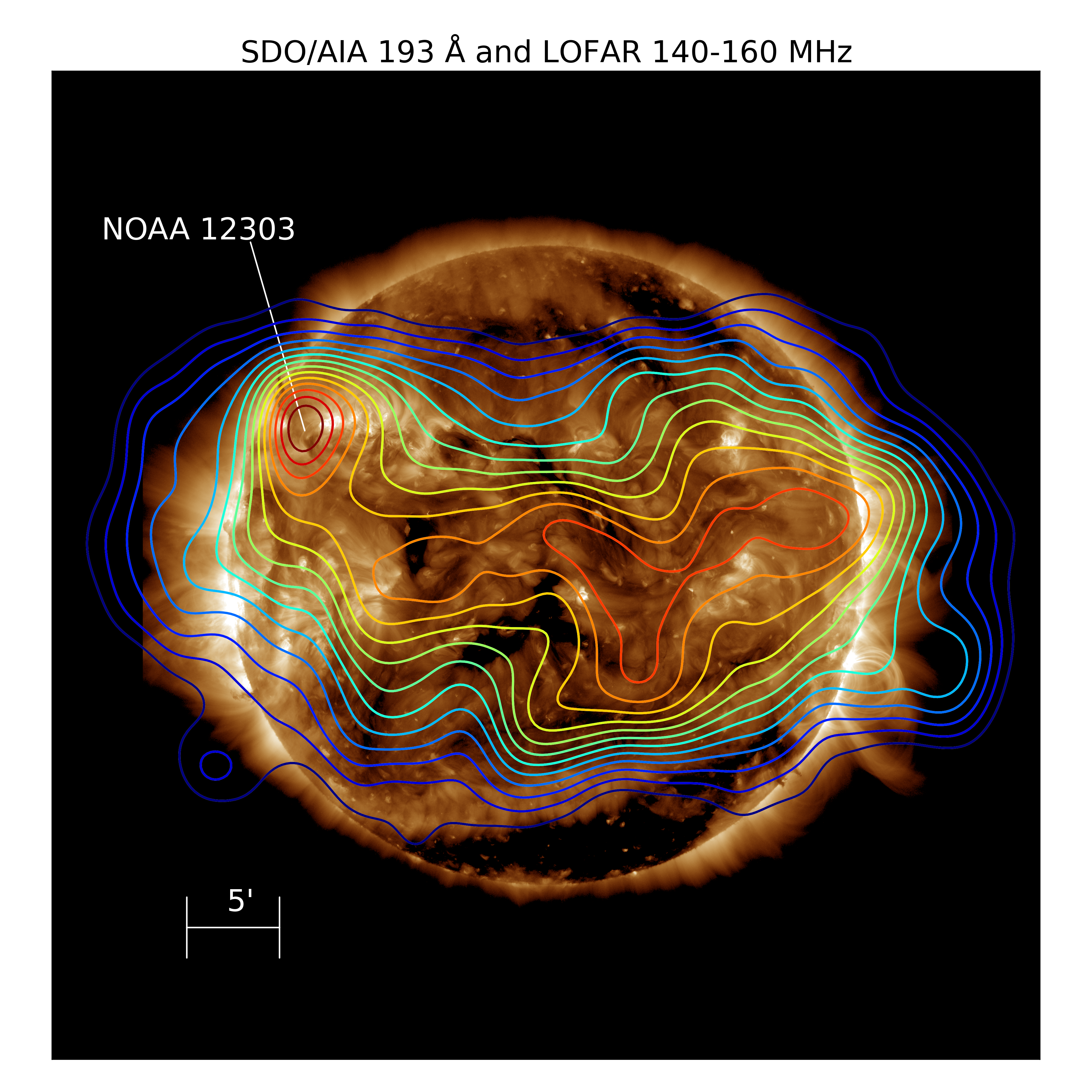}
  \caption{LOFAR contours 50-95$\%$ (blue to red) of the peak intensity on top of a SDO/AIA 193 \AA\space image. The contours are from a multi-frequency LOFAR map (140--160 MHz) and the 193 \AA\space EUV image is from 11:05 UT, 2015 March 20.}
  \label{fig:figure_1}
\end{figure}

On 2015 March 20 between 08:32 and 10:50 UT a partial solar eclipse (80 $\%$ totality) was observed as part of a 5-hour LOFAR observing campaign that was carried out between 07:20 and 12:00 UT. During the observation, there were a number of active regions (NOAA 12297, 12299, 12302, 12303, 12304) visible on the solar disc. There were also a number of C-class flares prior to the eclipse, most notably a C7.9 (which peaked at 00:58 UT) accompanied by a partial halo CME (first observed in SOHO's Large Angle and Spectrometric Coronagraph C2 \citep[LASCO;][]{brueckner1995large} field of view at 01:05 UT). A 193 \AA\space Solar Dynamics Observatory/Atmospheric Imaging Assembly (SDO; \cite{pesnell2011solar}, AIA; \cite{lemen2011atmospheric}) image of the Sun with LOFAR contours is shown in \autoref{fig:figure_1}.

LOFAR is a low-frequency radio interferometer operated by the Netherlands Institute for Radio Astronomy (ASTRON). LOFAR is comprised of thousands of antenna divided up into core, remote and international stations, centred around Exloo in the Netherlands and extending over a maximum baseline of $\sim$2000\,km. Each station is composed of Low Band Antennas (LBAs), observing from 10--90 MHz, and High Band Antennas (HBAs), which observe from 120--240 MHz.

In this study, the HBAs from 23 of LOFAR’s core stations were used, providing a maximum baseline of $\sim$3.5 km. Raw visibility data were produced using LOFAR’s interferometric mode for 253 baselines, providing a temporal resolution of 1s and spectral resolution of 12.207\,kHz \citep{van2013lofar}. Observations were taken for a number of subbands, every 10 MHz between 120 MHz and 180 MHz, and integrated over 5.5 seconds in order to increase signal-to-noise. The angular resolution of the 23 station array ranges from 2.0\arcmin\space at 120 MHz and 1.2\arcmin\space at 180 MHz. In the following analysis, the data from the core stations were used to produce interferometric maps and carry out the lunar de-occultation technique.

%%%%%%%%%%%%%%%%%%%%%%%%%%%%%%%%%%%%%%%%%%%%%%%%%%%%%%%%%%%%%%%%%%%%%%%%%%%%%%%%%%%%%%%%%
%							  DATA ANALYSIS 
%%%%%%%%%%%%%%%%%%%%%%%%%%%%%%%%%%%%%%%%%%%%%%%%%%%%%%%%%%%%%%%%%%%%%%%%%%%%%%%%%%%%%%%%%
\section{Data analysis}\label{sec:data_analysis}

% %%%%%%%%%%%%%%%%%%%%%%%%%%%%%%%%%%%%%%%%%%%%%%%%%%%%%%%%%%%%%%%%%%%%%%%%%%%%%%%%%%%%%%%%%
% %							INTERFEROMETRIC IMAGING 
% %%%%%%%%%%%%%%%%%%%%%%%%%%%%%%%%%%%%%%%%%%%%%%%%%%%%%%%%%%%%%%%%%%%%%%%%%%%%%%%%%%%%%%%%%
\subsection{Interferometric imaging}\label{ssec:interferometric_imaging}
\begin{figure}
  \centering
  \includegraphics[width=\columnwidth]{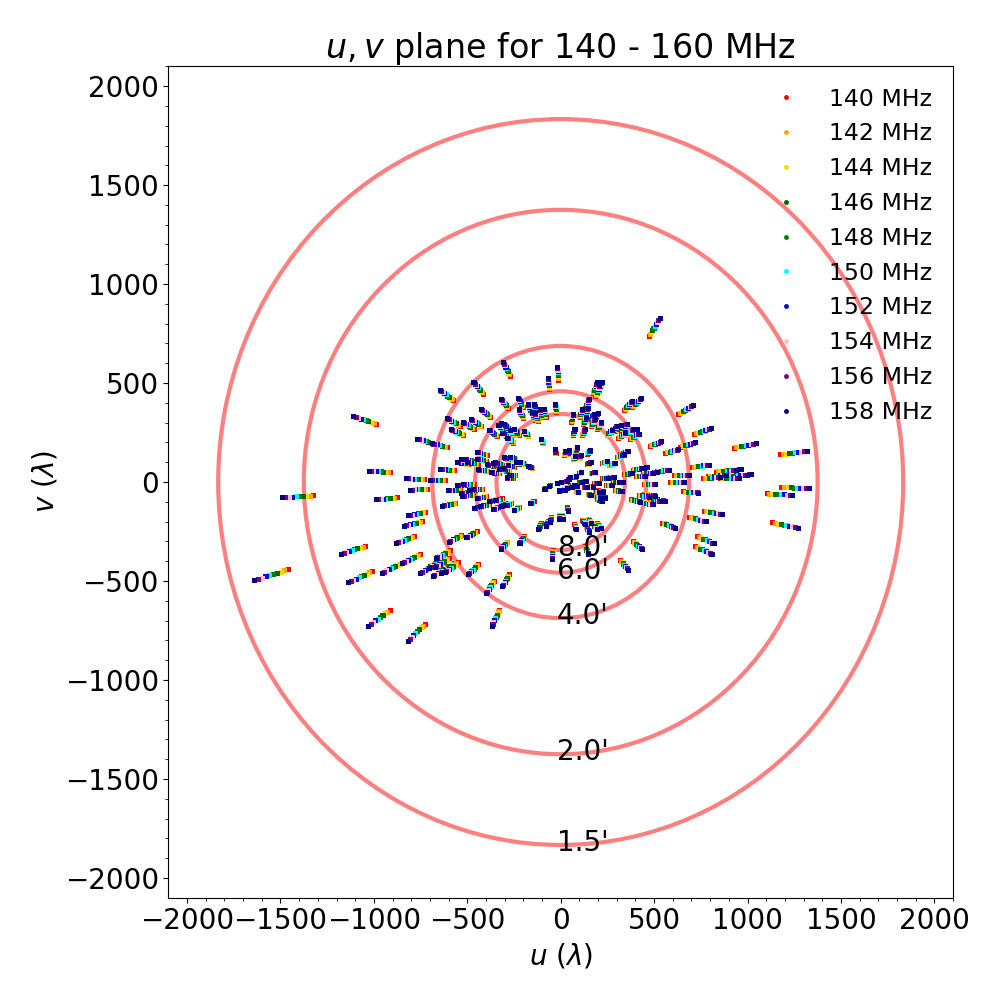}
  \caption{$u,v$ coverage for the multi-frequency band 140–160 MHz. Each of the coloured points on the $u,v$ plane correspond to a different frequency observed for a given baseline. A series of red rings illustrate an example of the baselines which contribute to achieving a particular angular resolution for 140 MHz, i.e. the longer baselines are responsible for resolving the smaller structure.}
  \label{fig:figure_2}
\end{figure}

\begin{figure*}[h!]
  \centering
  \includegraphics[width=\textwidth, trim=0cm 1cm 0cm 3cm]{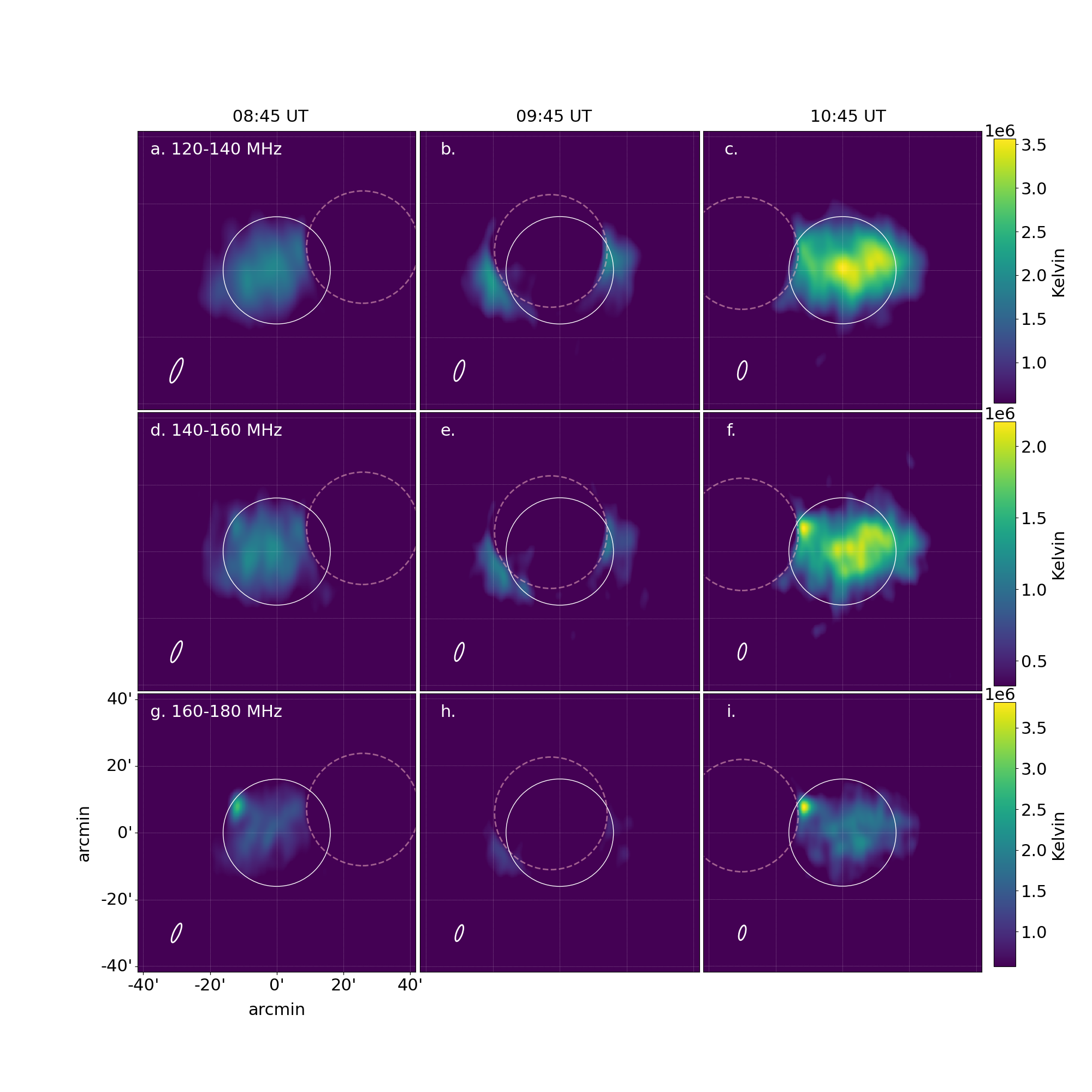}
  \caption{LOFAR multi-frequency maps at ingress, maximum phase, and egress of the partial solar eclipse on 2015 March 20. Time increases from left to right, whereby the first column is 08:45 UT, the second column is 09:45 UT, and the third column is 10:45 UT. Frequency increases from top to bottom of the grid. The top row is the 120--140 MHz multi-frequency band, the middle row is the 140--160 MHz multi-frequency band, and the bottom row is the 160--180 MHz multi-frequency-band. In each image, the solid white circle is the visible solar limb and the pink dashed outline is the lunar limb. The white ellipse in the bottom left corner of each image is representative of the beamsize.}
  \label{fig:figure_3}
\end{figure*}

This analysis was divided up into two distinct parts, namely the interferometric imaging of the solar eclipse and the implementation of the lunar de-occultation technique. Firstly, the Default Pre-Processing Pipeline \citep[DPPP;][]{2018ascl.soft04003V} was used to average the data over 2.5 seconds and apply weights to the visibilities using autocorrelations to account for the array configuration. DPPP is also capable of performing flagging for radio frequency interference (RFI). This was not applied as automated RFI flaggers are susceptible to flagging solar radio bursts and sometimes quiet solar emission. Instead, the data were manually inspected and flagged for RFI and malfunctioning antennas. A bandpass calibration was applied using static calibration tables generated from a 20 minute observation of Cygnus A prior to the eclipse. 

As this was an eclipse observation, longer exposure aperture synthesis was considered unsuitable. Snapshot imaging was used instead and in order to increase the $u,v$ coverage, multi-frequency synthesis (MFS) was implemented \citep{mccready1947solar, conway1990multi, 1999ASPC..180..419S}. The visibilities at several frequencies were concatenated into a multi-frequency band. A number of iterations of self-calibration were then applied using a multiscale, multi-frequency synthesis (MS-MFS) CLEAN \citep{mcmullin2007casa, rau2011multi}. MFS is a useful technique to increase $u,v$ coverage without the inclusion of additional baselines. \autoref{fig:figure_2} demonstrates that for a particular baseline it is possible to have many $u,v$ points corresponding to different frequencies. For this work a frequency range of 120–180 MHz was divided into three separate multi-frequency bands; 120–140 MHz, 140–160 MHz, and 160–180 MHz. Each band has a width of 20 MHz as the spectral brightness of the radio non-flaring Sun is known not to vary greatly over this range at these frequencies. 

In \autoref{fig:figure_1}, the contours of the solar radio emission observed by LOFAR were overlaid onto an SDO/AIA image from 11:05 UT. The LOFAR contours are shown to interweave between the coronal holes and small bright regions on the Sun as seen in the 193 \AA\space EUV image. The brightest source of radio emission observed, is situated close to the north-eastern limb, and is associated with the active region NOAA 12303. 

A series of multi-frequency CLEAN maps were produced every 10 minutes for the whole duration of the observation, an example of which can be seen in \autoref{fig:figure_3}. These maps are plotted in a helioprojective coordinate system. Each row is a different multi-frequency band, increasing in frequency from top to bottom. The first column is at 08:45 UT, the second column at 09:45 UT, and the third column at 10:45 UT. These CLEAN maps show clearly the passage of the Moon (pink dashed circle) across the Sun as the eclipse transitions through ingress, maximum phase (80 $\%$), and egress. 

Certain features are more apparent in the different multi-frequency maps as is shown in \autoref{fig:figure_3}. The structure in the lower frequency band images appears more diffuse, most likely due to the decreased angular resolution at these frequencies. This is evident in Figure 3 (c), (e), and (f) where the features appear extended in the lower frequency bands in comparison to the compact features seen in the higher frequency bands. The unique set-up of a solar eclipse allows for the implementation of the lunar de-occultation technique which can provide better spatial resolution than what is achieved via standard interferometric imaging (STIM) in \autoref{fig:figure_3}.

% %%%%%%%%%%%%%%%%%%%%%%%%%%%%%%%%%%%%%%%%%%%%%%%%%%%%%%%%%%%%%%%%%%%%%%%%%%%%%%%%%%%%%%%%%
% %							LUNAR DE-OCCULTATION
% %%%%%%%%%%%%%%%%%%%%%%%%%%%%%%%%%%%%%%%%%%%%%%%%%%%%%%%%%%%%%%%%%%%%%%%%%%%%%%%%%%%%%%%%%

\subsection{Lunar de-occultation}\label{ssec:lunar_deoccultation}
Lunar de-occultation is a unique method that exploits imaging observations of a solar eclipse. Over time, as the moon reveals the solar surface, the intensity in the maps changes due to the revelation of coronal structure. The change in intensity over time can be related to intensity variation in space. For this to work, it must be assumed that all changes in intensity are due to the de-occultation of coronal sources. To ensure that this was the case, we searched for various signs of activity during the de-ocultation. Firstly, we examined GOES X-ray lightcurves and radio dynamic spectra, both of which were found to be clear of significant solar activity. Any timesteps that were associated with small bursts in the dynamic spectrum were flagged and removed. Secondly, any instrumental effects not removed by the calibration were taken into account by normalising the maps. This was done by dividing the entire map by the average intensity in a quiet region for each timestep.

In practice, lunar de-occultation is carried out by subtracting data from consecutive timestamps during the egress of the eclipse and summing the differenced data together. Here we explore two different approaches to lunar de-occultation; namely image-differencing (IMD) and visibility-differencing (VISD). One iteration of both the IMD and VISD methods is shown in \autoref{fig:figure_4}(a) and (b) respectively. The following sections focus on the application of the lunar de-occultation techniques to the central multi-frequency band, 140--160 MHz. The central band was chosen as the lower and upper bands were dominated by significant radio frequency interference (RFI).

\subsubsection{Image-differencing (IMD)}\label{sssec:image-differencing}
\begin{figure}
  \centering
  \includegraphics[width=\columnwidth]{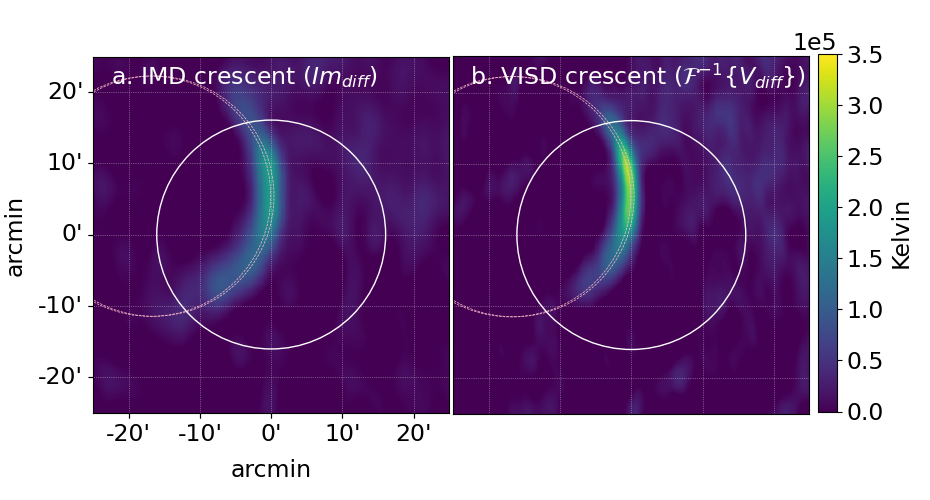}
  \caption{Demonstration of two different approaches to the lunar de-occultation technique. (a) contains the results from differencing two CLEAN maps via the image-differencing (IMD) method, i.e. $Im_{diff}$. (b) is the output when the visibility data are differenced first before CLEANing, as per the visibility-differencing (VISD) method, i.e. $F^{-1}\{V_{diff}\}$. In both (a) and (b), the differenced data are 1 minute apart. The solid white circle is the visible solar limb and the dashed pink circle is the visible lunar limb.}
  \label{fig:figure_4}
\end{figure}

The resulting map after one step of the image-differencing (IMD) method is shown in \autoref{fig:figure_4}(a). The IMD method involved firstly performing the inverse Fourier transform of visibility data to make a series of dirty maps with 1 minute cadence. These dirty maps were then CLEANed and self-calibrated using a MS-MSF CLEAN, as described above. The pixel values in each CLEAN map were subtracted from the pixel values of the following map, that is, $Im_{t+1}(x,y)$ - $Im_t(x,y)$, hereafter $Im(x,y)$ is written as $Im$. An example of a resulting differenced map, $Im_{diff}$, when two CLEAN maps 1 minute apart were differenced is shown in \autoref{fig:figure_4}(a). $Im_{diff}$ is a crescent-shaped portion of the solar corona. It is effectively an annular aperture which can be used to probe the coronal structure revealed in the time interval between $Im_{t+1}$ and $Im_t$. 

\subsubsection{Visibility-differencing (VISD)}
The resulting map after one step of the second approach, the visibility-differencing (VISD) method, is shown in \autoref{fig:figure_4}(b). With the VISD method, the raw visibility data were used instead. Consecutive visibility data were subtracted from each other, that is, $V_{t+1}(u,v)$ - $V_{t}(u,v)$, hereafter $V(u,v)$ is written as $V$. The inverse Fourier transform was then taken of differenced visibility data, $V_{diff}$, to produce a dirty map. The dirty maps were CLEANed and self-calibrated as mentioned previously, to produce $Im_{diff}$. An example of the CLEANed crescent-shaped annulus produced via VISD can be seen in \autoref{fig:figure_4}(b) whereby visibility data 1 minute apart were differenced. 

The last step taken in both the IMD and VISD methods was to produce a map comparable to an image made via standard interferometric imaging (STIM) at 10:59 UT. As described by \cite{gary1987multifrequency}, the Point Spread Function (PSF) is modified when using this de-occultation technique. In 1 minute the 'knife-edge' lunar limb profile has moved by 0.6\arcmin. This is convolved with the instrument beam producing a ramp-like profile. When two of these ramps are differenced it results in a triangular-shaped profile. This causes the emission extending beyond the dashed pink circle in \autoref{fig:figure_4}(a). The convolution of the instrument beam and the triangular window results in a broader Gaussian beam with a lower amplitude than the synthesised beam. This was taken into account by dividing each $Im_{diff}$ CLEAN map by a correction factor of 0.1, calculated using Equation 1 of \cite{gary1987multifrequency}.

\begin{figure}[h!]
  \centering
  \includegraphics[width=0.83\columnwidth, trim = 0 0 0 0.5 cm]{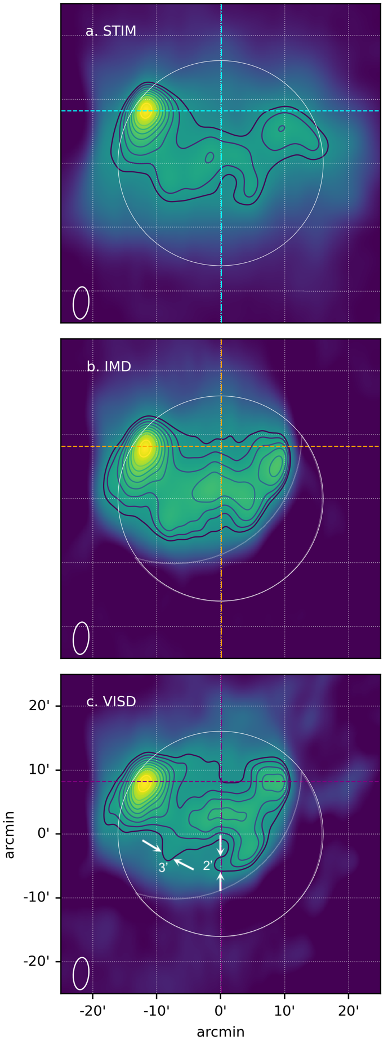}
  \caption{Three LOFAR maps produced using three different methods. (a) A CLEAN map made via standard interferometric imaging (STIM) at 10:59~UT. (b) A map made by summing the $Im_{diff}$ produced by differencing consecutive CLEAN maps, 1 minute apart (IMD). (c) A map made by summing the $Im_{diff}$, made by differencing consecutive visibility data, 1 minute apart (VISD). Each map is normalised and the contours are 50--95~$\%$ the max intensity. The solid white circle is the visible solar limb. The white crescent is the area not de-occulted by the moon.\vspace{-8 mm}}
  \label{fig:figure_5}
\end{figure}

After this correction was applied, the differenced maps were summed together, that is, $\Sigma Im_{diff}[t]$. \autoref{fig:figure_5} is a comparison of a map made using STIM and maps made using the IMD and VISD methods. It is clear from \autoref{fig:figure_5}(b) and (c) that finer structure in the quiet Sun is revealed when using the lunar de-occultation technique. This observation of smaller sources in the corona implies that the effects of scattering are not as severe as it might be concluded from \autoref{fig:figure_5}(a), where radio sources appear larger and broader. In order to quantify the obtained improved resolution, 1D intensity profiles were taken across the brightest emission (corresponding with the active region NOAA 12303) and the quiet Sun in each map. 

% %%%%%%%%%%%%%%%%%%%%%%%%%%%%%%%%%%%%%%%%%%%%%%%%%%%%%%%%%%%%%%%%%%%%%%%%%%%%%%%%%%%%%%%%%
% %							SOURCE SIZE DETERMINATION
% %%%%%%%%%%%%%%%%%%%%%%%%%%%%%%%%%%%%%%%%%%%%%%%%%%%%%%%%%%%%%%%%%%%%%%%%%%%%%%%%%%%%%%%%%

\subsection{Source size determination}\label{ssec:source_size_determination}
In order to directly compare the maps shown in \autoref{fig:figure_5}, their intensities were first normalised. One of the brighter sources in all three maps is associated with the active region NOAA 12303. A horizontal slice was taken across the bright emission in each map, depicted by a cyan, orange or purple dashed line in \autoref{fig:figure_5}(a)--(c). Another vertical slice was taken across a region of quiet Sun in each map, represented my the dotted-dashed coloured lines in \autoref{fig:figure_5}(a)--(c). The width of the highest peak in each intensity profile was measured at 80 $\%$ the maxiumum intensity, the results of which are detailed in \autoref{table:table_1}.

% %%%%%%%%%%%%%%%%%%%%%%%%%%%%%%%%%%%%%%%%%%%%%%%%%%%%%%%%%%%%%%%%%%%%%%%%%%%%%%%%%%%%%%%%%
% %							RESULTS
% %%%%%%%%%%%%%%%%%%%%%%%%%%%%%%%%%%%%%%%%%%%%%%%%%%%%%%%%%%%%%%%%%%%%%%%%%%%%%%%%%%%%%%%%%

\section{Results}\label{sec:results} 
\subsection{Comparison of imaging techniques}
\begin{figure}[h!]
  \centering
  \includegraphics[width=\columnwidth]{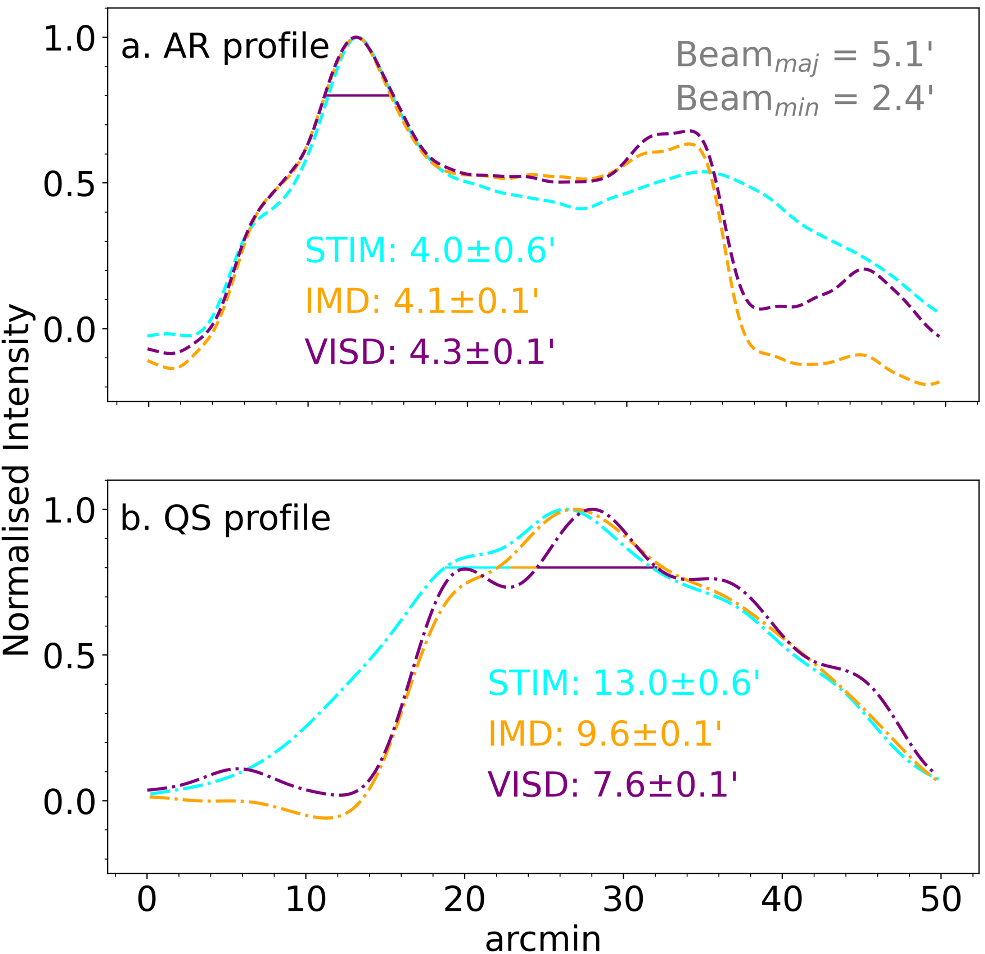}
  \caption{Plot of the intensity slices across the bright active region (AR) emission corresponding to NOAA 12303 (- -) and the quiet Sun (QS) emission (-.). The profile colours correspond to the horizontal and vertical, coloured, dashed lines in \autoref{fig:figure_5}(a)--(c). The widths have been determined for the tallest peak, in each profile, as is labelled above.}  
  \label{fig:figure_8}
\end{figure}

From \autoref{fig:figure_5}, it is apparent that the maps from both de-occultation techniques display more detail than those when using STIM. 1D slices were taken across the radio emission associated with active region NOAA 12303 (AR profile) and a region of quiet Sun (QS profile) in each of the STIM, IMD, and VISD maps. The intensity profiles in \autoref{fig:figure_8} correspond to the same coloured lines in \autoref{fig:figure_5}. The width of the tallest peaks in each intensity profile can be seen in \autoref{fig:figure_8} and is also detailed in \autoref{table:table_1}. The error in the width for the STIM map was taken to be 1/4 of the beamsize at this frequency (0.6\arcmin). The error in the widths for both the IMD and VISD maps was calculated as 1/4 of the crescent aperture de-occulted in 1 minute (0.1\arcmin). 

\begin{table}[h!]
\begin{tabular}{l|l|l|l}
          & STIM & IMD & VISD \\ \hline
$\theta$ & 2.4$\pm$0.6\arcmin & 0.6$\pm$0.1\arcmin & 0.6$\pm$0.1\arcmin \\
AR width & 4.0$\pm$0.6\arcmin            & 4.1$\pm$0.1\arcmin            & 4.3$\pm$0.1\arcmin    \\ 
QS width & 13.0$\pm$0.6 & 9.6$\pm$0.1\arcmin & 7.6$\pm$0.1\arcmin 
\end{tabular}
\caption{Width of sources at 80 $\%$ maximum intensity in the maps produced via the three imaging techniques; standard interferometric imaging (STIM), image-differencing (IMD), and visibility differencing (VISD). The beamwidth ($\theta$) for STIM is given in the  direction of the minor axis.}
\label{table:table_1}
\end{table}

From \autoref{table:table_1}, the peak in the AR profile was found to be 4.3$\pm$0.6\arcmin\space in the VISD map, 4.1$\pm$0.1\arcmin\space in the IMD map, and 4.0$\pm$0.1\arcmin\space in the STIM map. These are all within error of each other and therefore no improved resolution was achieved in the AR. However, it is clear from \autoref{fig:figure_5} that the contours around the AR in both the VISD and IMD maps suggest a more complex morphology that is not resolved by the STIM map.

In addition, from examination of \autoref{fig:figure_5}(a)--(c), it is evident that there is also finer structure resolved in the QS by the lunar de-occultation techniques. The contours marked by white arrows in \autoref{fig:figure_5}(c) provide tentative evidence of a structure as small as 2--3\arcmin\space that is not resolved by the STIM map of \autoref{fig:figure_5}(a). In addition, the VISD profile in \autoref{fig:figure_8}(b) appears to reveal a double-peaked feature more clearly instead of the single-peaked feature of the STIM and IMD profiles. The width of the QS source was found to be 7.6$\pm$0.1\arcmin\space in the VISD map, as shown in \autoref{table:table_1}. This is 1.4 times smaller than that of the IMD profile (9.6$\pm$0.1\arcmin) and 1.7 times smaller than that of the STIM profile (13.0$\pm$0.6\arcmin). This implies that this QS source size is overestimated by 40--70 $\%$ when using the IMD lunar de-occultation technique or STIM. This has implications for determining the effects of scattering, that is, if QS source sizes from STIM or IMD were used to determine the level of scattering, the effect would be overestimated. The following section seeks to quantify this overestimation.

 %%%%%%%%%%%%%%%%%%%%%%%%%%%%%%%%%%%%%%%%%%%%%%%%%%%%%%%%%%%%%%%%%%%%%%%%%%%%%%%%%%%%%%%%%
 %							DISCUSSION
 %%%%%%%%%%%%%%%%%%%%%%%%%%%%%%%%%%%%%%%%%%%%%%%%%%%%%%%%%%%%%%%%%%%%%%%%%%%%%%%%%%%%%%%%%
\section{Discussion}\label{sec:discussion}
Theoretically, the IMD and VISD methods offer an angular resolution that is a factor of $\sim$4 times better than that of the STIM method. In this study, the VISD and IMD methods reveal a QS source to be $\sim$7.6\arcmin\space and 9.6\arcmin\space in width, which are 1.7 and 1.4 times smaller than the sources revealed by the STIM method. As discussed in \autoref{sec:results}, finer structure is evident in both the IMD and VISD maps of the quiet Sun. However, though smaller scales are found in the QS source, all three methods yield similar results for the AR source, which was found to be $\sim$4\arcmin. This difference may be a consequence of radio wave scattering being larger in the active region. That is, the plasma in and above that AR is expected to have more density fluctuations than the QS due to higher levels of turbulence \citep{abramenko2010intermittency, abramenko2020analysis}. Increased levels of turbulence results in more scattering and larger source sizes.

Both \autoref{fig:figure_5}(b) and (c) reveal structure that is not obvious in \autoref{fig:figure_5}(a) around both the AR and QS sources. The resolution of smaller structure is due to the data differencing implemented during the lunar de-occultation procedure. Differencing consecutive visibilities or maps results in the removal of the background as well as enhanced spatial resolution. In this case the background is in fact the QS emission. By removing it, one reveals the small-scale structure normally embedded in the QS background. Using a similar approach \citet{marsh1980vla} were able to resolve source sizes of 9-25\arcsec\space at 4.9 GHz. Here, the smallest source size resolved using IMD and VISD was the AR, measuring 4.1$\pm$0.1\arcmin and 4.3$\pm$0.1\arcmin as well as tentative evidence of features as small as 2--3~\arcmin.

The difference in resolution achieved by the two lunar de-occultation approaches (IMD and VISD) may be attributed to the different levels of noise in each of the approaches. By making a series of maps, differencing them and then making a CLEAN map, artefacts are introduced via the imaging procedure for each image in the series, that is, $Im_{t}$, $Im_{t+1}$... $Im_{n}$. However, by differencing the visibilities first and then making a CLEAN map, the introduction of artefacts via the imaging procedure is only once at the final step, that is, making $Im_{diff}$. 

The overestimation of source sizes by STIM in the QS corona has implications for our approximation of the effects of scattering as these two properties are intrinsically related. Using equations (3), (6), and (7) from \cite{steinberg1971coronal} and equation (8) from \cite{chrysaphi2018cme} angular broadening, $d\langle\theta^2\rangle/dr$, can be related to the level of turbulence caused by coronal density inhomogeneities, $\delta n^2/n^2$.
\begin{equation}
\centering
    \frac{d\langle\theta^2\rangle}{dr} = \frac{\sqrt{\pi}}{2h}\frac{f_{pe}^4(r)}{(f^2 - f_{pe}^2(r))^2}\frac{\langle\delta n^2\rangle}{n^2}
    \label{eqn:ang_brodening_scattering}
\end{equation}
where $h$ is the correlation scale for the inhomogeneities in the corona, $f_{pe}$ is the plasma frequency, and $f$ the frequency of the observed emission. Equation 1 shows angular broadening is proportional to the relative level of density fluctuation. Therefore, by taking the ratio of angular broadening in the QS between VISD and STIM, scattering is overestimated by a factor of 70~$\%$. Similarly, IMD would yield an overestimation in the effects of scattering in the QS by a factor of 40 $\%$. This suggests a possible over-estimation of the effects of radio scattering in the corona in previous studies \citep{kontar2017imaging}.  
 
 %%%%%%%%%%%%%%%%%%%%%%%%%%%%%%%%%%%%%%%%%%%%%%%%%%%%%%%%%%%%%%%%%%%%%%%%%%%%%%%%%%%%%%%%%
 %						    CONCLUSIONS
 %%%%%%%%%%%%%%%%%%%%%%%%%%%%%%%%%%%%%%%%%%%%%%%%%%%%%%%%%%%%%%%%%%%%%%%%%%%%%%%%%%%%%%%%%
\section{Conclusions}\label{sec:conclusions}
In summary, due to propagation effects such as refraction and scattering, it is accepted that observed sources in the solar corona may undergo changes in shape, position, and indeed size. Presently, the actual extent of coronal scattering is still not fully understood. This work has highlighted the possible issues associated with using observed source size in radio imaging to constrain the effects of scattering. 

The ambiguity surrounding the effects of scattering has been abetted by, up until now, the rarity of high spatial resolution imaging of the Sun. Therefore, it was impossible to understand if the lack of observations of the sub-arcminute source sizes is due to scattering effects or, due to insufficient resolution of the observations.

Here the first LOFAR observation of a solar eclipse was presented. This rare observation has provided a unique opportunity to probe coronal source sizes and push interferometric imaging beyond its limit when longer baseline observations were not available. It enabled the use of a special technique, namely lunar de-occultation, to achieve higher spatial resolution than that attainable via standard interferometric imaging. Using the VISD lunar de-occultation technique source sizes as small as a few arcminutes were resolved. This agrees with previous studies that claim the effects of scattering at low frequencies prevent sub-arcminute structure being observed regardless of the angular resolution of the instrument \citep{bastian1994angular}.  

Due to increased turbulence around the AR \citep{abramenko2010intermittency, abramenko2020analysis} the effects of scattering were deemed so severe that, regardless of the imaging technique, no better resolution was achieved. However, the lunar de-occultation techniques provided better resolution than standard interferometric imaging of the QS corona. Similar results were found at microwave frequencies by \cite{marsh1980vla} and \cite{gary1987multifrequency}. A difference in results depending on which approach was taken when performing the lunar de-occultation technique was noted. The smallest source sizes were found in the maps made via the VISD method. An over-estimation of QS source sizes by a factor of 1.4--1.7 when using IMD or STIM was demonstrated, highlighting the implications for estimation of the effects of coronal scattering. As angular broadening and coronal turbulence are directly proportional, QS sources measured in maps from IMD or STIM maps would yield an over-estimation of scattering of 40--70~$\%$. 

This work endorses the use of longer baseline solar imaging, to push the limits of high spatial resolution interferometers in order to more accurately quantify the effects of scattering. Though solar eclipses are infrequent events, the addition of a number of low-frequency radio interferometers around the world (such as the MWA) increases our chance of performing similar analysis at even lower frequencies where the effects of radio wave propagation are known to be even more severe.

 %%%%%%%%%%%%%%%%%%%%%%%%%%%%%%%%%%%%%%%%%%%%%%%%%%%%%%%%%%%%%%%%%%%%%%%%%%%%%%%%%%%%%%%%%
 %							ACKNOWLEDGEMENTS
 %%%%%%%%%%%%%%%%%%%%%%%%%%%%%%%%%%%%%%%%%%%%%%%%%%%%%%%%%%%%%%%%%%%%%%%%%%%%%%%%%%%%%%%%%
\begin{acknowledgements}
A. M. Ryan's research is jointly funded by the Irish Research Council and AstroTec Holding B.V. as part of the Irish Research Council Enterprise Partnership Scheme and hosted by the Dublin Institute for Advanced Studies (DIAS). E. P. Carley is supported by the Schrödinger Fellowship at DIAS. This paper is based on data obtained with the International LOFAR Telescope (ILT) under project code DDT 003. LOFAR \citep{van2013lofar} is the Low Frequency Array designed and constructed by ASTRON. It has observing, data processing, and data storage facilities in several countries, that are owned by various parties (each with their own funding sources), and that are collectively operated by the ILT foundation under a joint scientific policy. The ILT resources have benefitted from the following recent major funding sources: CNRS-INSU, Observatoire de Paris and Université d'Orléans, France; BMBF, MIWF-NRW, MPG, Germany; Science Foundation Ireland (SFI), Department of Business, Enterprise and Innovation (DBEI), Ireland; NWO, The Netherlands; The Science and Technology Facilities Council, UK; Ministry of Science and Higher Education, Poland.

\newline A. M. Ryan would like to pay special thanks to Eduard Kontar, Tim Bastian, Dale Gary for discussions of coronal scattering and lunar de-occultation techniques.

\end{acknowledgements}

 %%%%%%%%%%%%%%%%%%%%%%%%%%%%%%%%%%%%%%%%%%%%%%%%%%%%%%%%%%%%%%%%%%%%%%%%%%%%%%%%%%%%%%%%%
 %							BIBLIOGRAPHY
 %%%%%%%%%%%%%%%%%%%%%%%%%%%%%%%%%%%%%%%%%%%%%%%%%%%%%%%%%%%%%%%%%%%%%%%%%%%%%%%%%%%%%%%%%
\bibliographystyle{aa}
\bibliography{bib.bib}

\end{document}